\documentstyle[epsfig]{aipproc}
\begin{document}
\title{Synchrotron Emission as the Source of GRB Spectra, Part II:
Observations.}

\author{Nicole M. Lloyd$^{1}$, Vah\'e Petrosian$^{1}$, and Robert D.
Preece$^{2}$}
\address{(1) Stanford University,
Stanford, California 94305\\
(2) University of Alabama in Huntsville, Huntsville, AL 35899}

\maketitle

\begin{abstract}
We test the models of synchrotron emission presented in Part I of
this series  (Lloyd \& Petrosian,
these proceedings; \cite{partI}) against the distributions and evolution of GRB spectral
parameters (particularly the low energy index, $\alpha$).  With knowledge
of the $E_{p}$ distribution and the correlation between $\alpha$ and
$E_{p}$ presented in \cite{partI}, we show how to derive the expected
distribution of $\alpha$ from fits to optically
thin synchrotron spectra, and compare this
with the observed distribution.  We show that
there is no difficulty explaining bursts below the
``line of death'', $\alpha < -2/3$, and that these bursts indicate
that the spectrum of accelerated electrons must flatten or
decline at low energies. Bursts with low energy 
spectral indices that fall
above this
limit are explained by the synchrotron self-absorption frequency
entering the lower end of the BATSE window.   Finally, we
 discuss a variety of spectral evolution behavior
seen in GRBs and explain this behavior in the context of synchrotron emission
from internal shocks.
\end{abstract}

\section*{Introduction}
The high and low energy spectral indices of GRB spectra contain
important information about GRB physics.  
The high energy photon index, $\beta$, usually 
reflects the steepness
of the underlying particle
energy distribution, in any non-thermal emission model.
The asymptotic value of the low energy photon index, $\alpha$, however,
 varies from model to model and can distinguish between  
different scenarios for GRB emission.  
 The peak and
dispersion in the distribution of the index $\alpha$ are difficult
to explain in the usual simple synchrotron model (SSM -
optically thin synchrotron emission from a power law
distribution of electrons with some minimum cutoff).
 Preece et al. \cite{pre98b} point out flaws in the SSM primarily based on a significant
fraction of bursts above the ``line of death'' value of $-2/3$.  Others \cite{lian96},
\cite{crid97} have pointed out
that the evolution of $\alpha$ throughout
the time history of the GRB is difficult to explain
by synchrotron emission in simple GRB emission models (i.e. external
shock models).  Indeed all of these phenomena must be explained by
any GRB emission mechanism.

In this paper, we show that synchrotron emission can accomodate both
the distribution and temporal
evolution of GRB spectral parameters.  We focus particularly
on the low energy spectral index, $\alpha$, because - again - synchrotron
models make definite predictions about the value of $\alpha$.
As shown in \cite{partI}, there is a strong correlation
between the value of $E_{p}$ and the value of
the ``asymptote'', $\alpha$, as determined by a Band function \cite{band93}
fit to the data from BATSE, limited to $25$ keV to
about $1.5$ MeV. We can use this relationship and knowledge
of the $E_{p}$ distribution to determine the resultant $\alpha$ distribution.
Finally, we give examples of spectral evolution in GRBs and show how this
is consistent with synchrotron emission from internal shocks.

\section*{The $\alpha$ Distribution}

  {As shown in \cite{partI}, there exists a relationship between the values
  of $\alpha$ and
  $E_{p}$  obtained from spectral fits 
  - the lower the value of $E_{p}$ (i.e. as at moves closer
  toward the low energy edge of the BATSE window), the lower (softer)
  the value of $\alpha$. Given
the mean and dispersion of the observed
$E_{p}$ distribution (\cite{pre99}), we can test if the peak
and dispersion
in the observed {$\alpha$ distribution} can
 be attributed to this correlation.}  
{We approximate the correlation
between $\alpha$ and $E_{p}$ by a simple analytical
function; log$(E_{p}) = h(\alpha)$ (the function $h(\alpha)$
depends on the specifics of the synchrotron model (see Figure 2 in \cite{partI})).}
{We then approximate the
$E_{p}$ distribution, $f({\rm log}(E_{p}))$ by a Gaussian in
 log$(E_{p})$, 
with a mean and dispersion representative of the
observed distribution. 
[It is important to point out that there has been considerable
controversy
over whether the observed distribution of $E_{p}$ is real or suffers from
selection bias \cite{lp1}, \cite{coh98}.  In the
past, we have estimated the selection bias in $E_{p}$ without accounting
for the non-diagonality of the detector response matrices (DRMs), which allow
for photons from higher energies (outside the BATSE band) to scatter
to low energies (into the BATSE band).  The DRMs reduce the selection bias,
but it is still not clear to what degree since there does not
exist a complete sample (in terms
of brightness) of bursts with spectral fits (see, e.g., \cite{lpm}).
As a result, we use the most conservative form of the $E_{p}$ distribution
in our analysis - the raw observed BATSE distribution.]}
{The distribution of $\alpha$ is then obtained from
the relation:
\begin{equation}
 g(\alpha) = f({\rm log}(E_{p}))\frac{dh(\alpha)}
 {d\alpha}.
\end{equation}}
Figure 2 
compares the resultant $\alpha$ distributions for
a sharp ($q=\infty$, right solid curve), intermediate ($q=2$, middle short-dashed
curve),  and flat ($q=0$, left long-dashed curve)
cutoff to the electron distribution with the observed
distribution obtained from fits to the BATSE data using the Band
spectral form.
\begin{figure}
\centerline{\epsfig{file=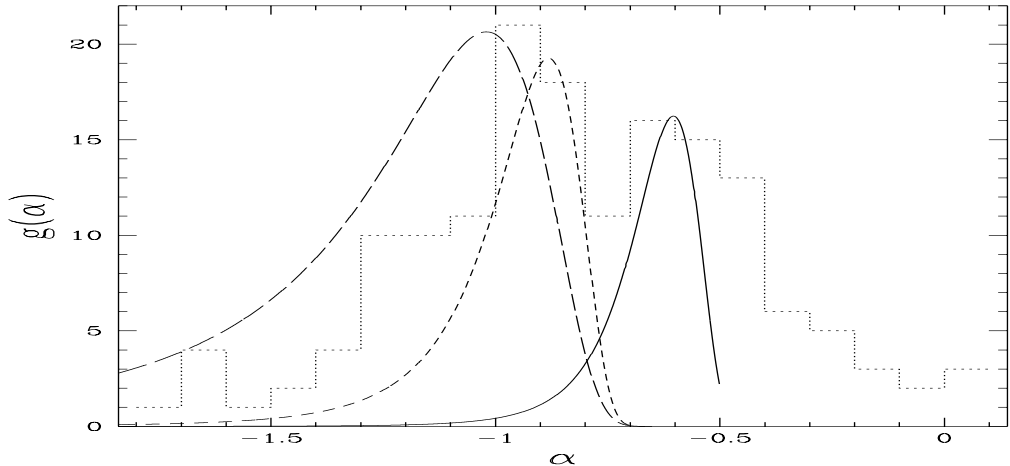
,width=.95\textwidth,height=0.45\textwidth}}
\caption{Simulated distributions of $\alpha$ for a sharp 
($q=\infty$, right solid curve), intermediate ($q=2$, middle short-dashed
curve),  and flat ($q=0$, left long-dashed curve)
 cutoff to the electron distribution, 
and the observed distribution (dotted histogram).}
\end{figure}
Several conclusions can be reached from this comparison. 
First, given a distribution in $q$, an instantaneous 
optically thin spectrum can easily accomodate bursts
with $\alpha < -2/3$ or below the line of death, where most
of the bursts are located.  
 The second conclusion is that the electron  energy distribution
below the turnover energy $E_{*}$ must be falling off, or at
least flat ($q \le 0$).  Otherwise, the SSM would predict
too many bursts with $\alpha$ less than about $-1.5$.  This
restriction will become stronger for a more realistic and 
broader distribution of $E_{p}$.

As, pointed out by
others \cite{pre98a}, the 
SSM fails to explain the
bursts with $\alpha > -2/3$, above the ``line of death''.
We believe these bursts that lie beyond our simulated distributions
may be physically explained by a self-absorption cutoff entering
the BATSE window (however, see also
\cite{gran}).  As shown in \cite{partI}, for some bursts synchrotron
self-absorption is necessary to provide a good spectral fit to the data.
However, we usually do not see the sharp $\alpha=3/2$ ($\nu_{min}<\nu \ll \nu_{abs}$)
or $\alpha=1.0$ ($\nu \ll {\rm min}[\nu_{min},\nu_{abs}]$) cutoff; the average value
of $\alpha$ above the line of death is about $0$.  There are two things that
could give a low $\alpha$ when the absorption cutoff is present in the BATSE
window.  One is due to the correlation discussed above (as $E_{p}$ moves closer
to the lower edge of the BATSE window, a lower (softer) value of $\alpha$ is measured).
But there is an additional factor when two breaks are present.  
If $\nu_{min}>\nu_{abs}$
and our fits place $E_{p} \propto \nu_{min}$, then
 when the self-absorption
frequency enters into the BATSE window, the Band spectrum cannot
accomodate this additional break.  As a result, the low energy
index ends up being a weighted average (depending on the relative
values of $\nu_{min}$ and $\nu_{abs}$ of the optically thin (-2/3)
and optically thick (1) asymptotes.  Support for this idea comes
from the GINGA data \cite{stroh98}; we believe their low $E_{p}$ values
not measured by BATSE are due to this absorption break $\nu_{abs}$ and
not $\nu_{min}$.
%

\section*{Spectral Evolution}
 GRB spectra are known to vary throughout the duration of the burst,
 and can even vary during individual pulses.  If the above
 explanation for the $\alpha$ distribution is correct, we would then
 expect a strong positive correlation in the time histories of
 $\alpha$ and $E_{p}$ for most bursts.  This relation should reflect the $h(\alpha)$
 curves shown in Figure 2 of \cite{partI} unless the parameters describing 
 the distribution of the accelerated particles 
 ($q$, $E_{*}$, $p$, in \cite{partI}) or the
 magnetic field vary during a pulse 
 or from pulse to pulse during a burst.  Depending
 on the nature of these variations, the expected correlation could
 be strengthened or weakened.
 
 There have been many studies of the time evolution
of spectral parameters (e.g. \cite{nor86}, \cite{ford95}, \cite{crid97}, \cite{pre98a}). 
%
{Crider et al. \cite{crid97} investigate the behavior of the low
energy spectral index $\alpha$ for a sample of 30 BATSE
GRBs.  They find that 18 of these bursts show
{hard-to-soft}
evolution of $\alpha$, while 12 exhibit {``tracking''} of the burst
time profile, $\alpha(t) \propto A(t)$, where
$A(t)$ is the amplitude of the photon spectrum.
  All of these bursts show a strong correlation 
between $\alpha$ and 
the peak energy, $E_{p}$, as a function of time.
  Recently, Preece et al. \cite{pre99}
published a catalog
of spectral data with high time resolution. 
 We have examined a sample of 46 bursts from this data set, and find
a variety of behaviors for $\alpha(t)$
and $E_{p}(t)$:
 (a) Both $\alpha$ and $E_{p}$ ``track'' the flux in time.
  This can be explained by the correlation between
$\alpha$ and $E_{p}$ discussed above (this, then, implies an intrinsic
correlation between $E_{p}$ and the flux throughout this burst). (b)
The parameter $\alpha$ tracks  the flux; $E_{p}$ varies on the
same timescale as the flux, but is in an envelope of hard-to-soft
evolution.  This can be explained by a sharpening of the cutoff
of the electron distribution from pulse to pulse.  Note the correlation
between $\alpha$ and $E_{p}$ is seen within each individual pulse. (c)
The parameter $\alpha$ appears to evolve from hard (above
the ``line of death'') to soft, while $E_{p}$ fluctuates around fairly
high ($\sim 500$ keV) values.  This can be explained by a transition
from an optically thick (self-absorption) to optically
thin regime throughout the duration of the burst.

All of the observed spectral behaviors we have encountered
be explianed by i) the correlation between
$\alpha$ and $E_{p}$, and ii) regarding each pulse as an independent emission
episode (as one would expect for internal
shocks, for example), which allows for evolution of the smoothness of the
particle energy distribution cutoff as well as the optical
depth in the shock. We will discuss this in more detail
in a future publication.

\begin{figure}
\centerline{\epsfig{file=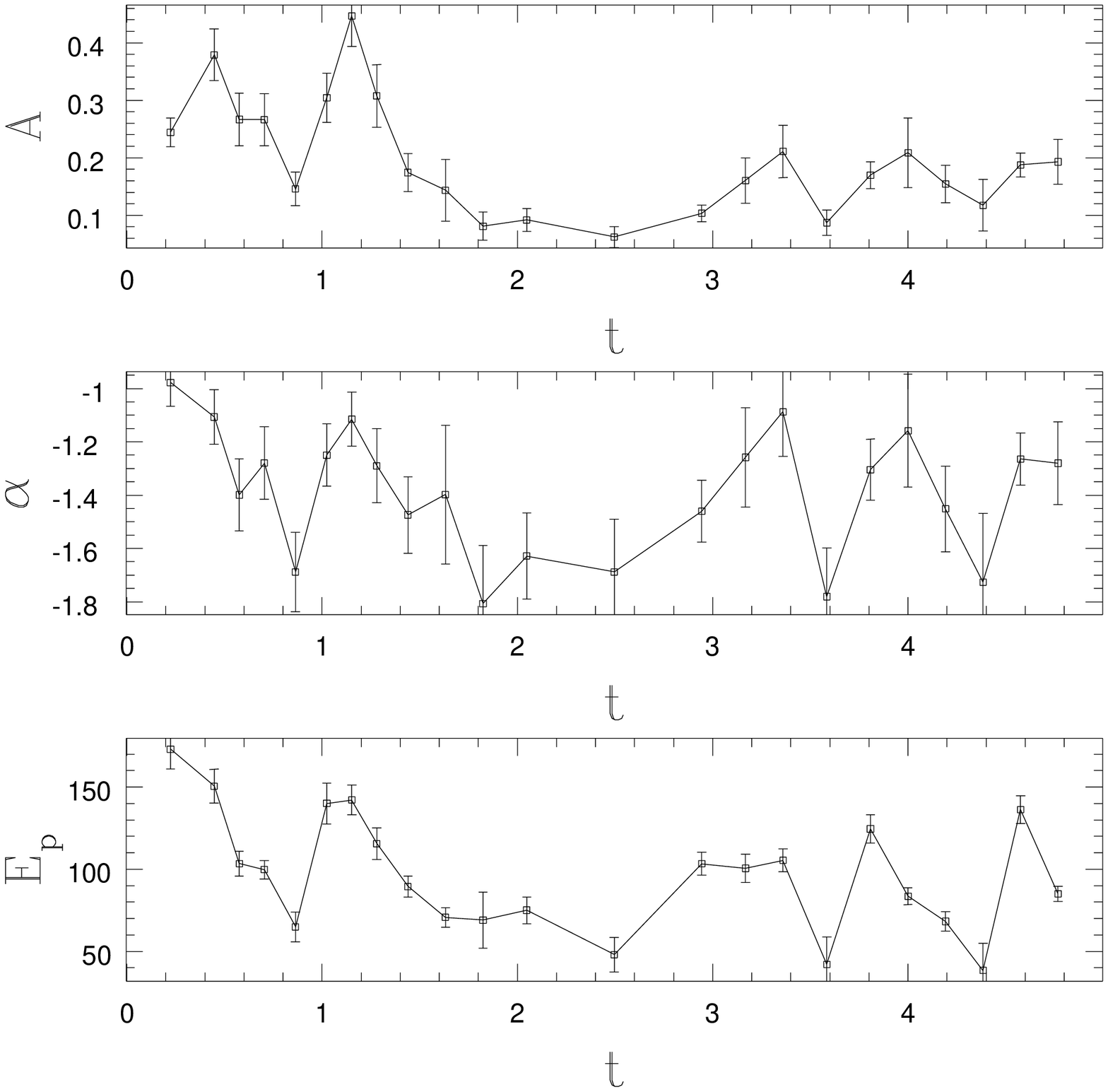,width=0.33\textwidth,height=0.4\textwidth}
\epsfig{file=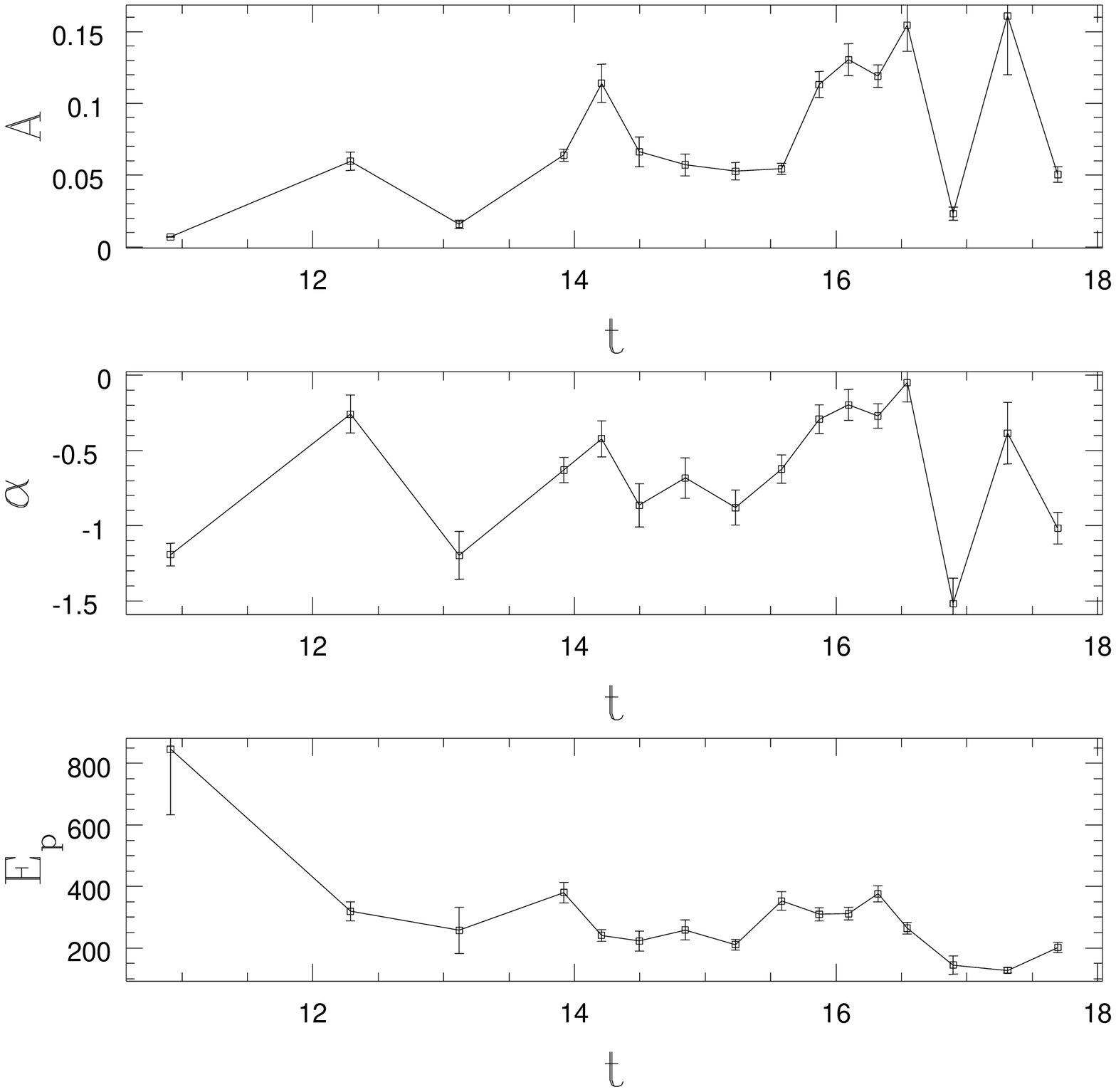,width=0.33\textwidth,height=0.4\textwidth} 
\epsfig{file=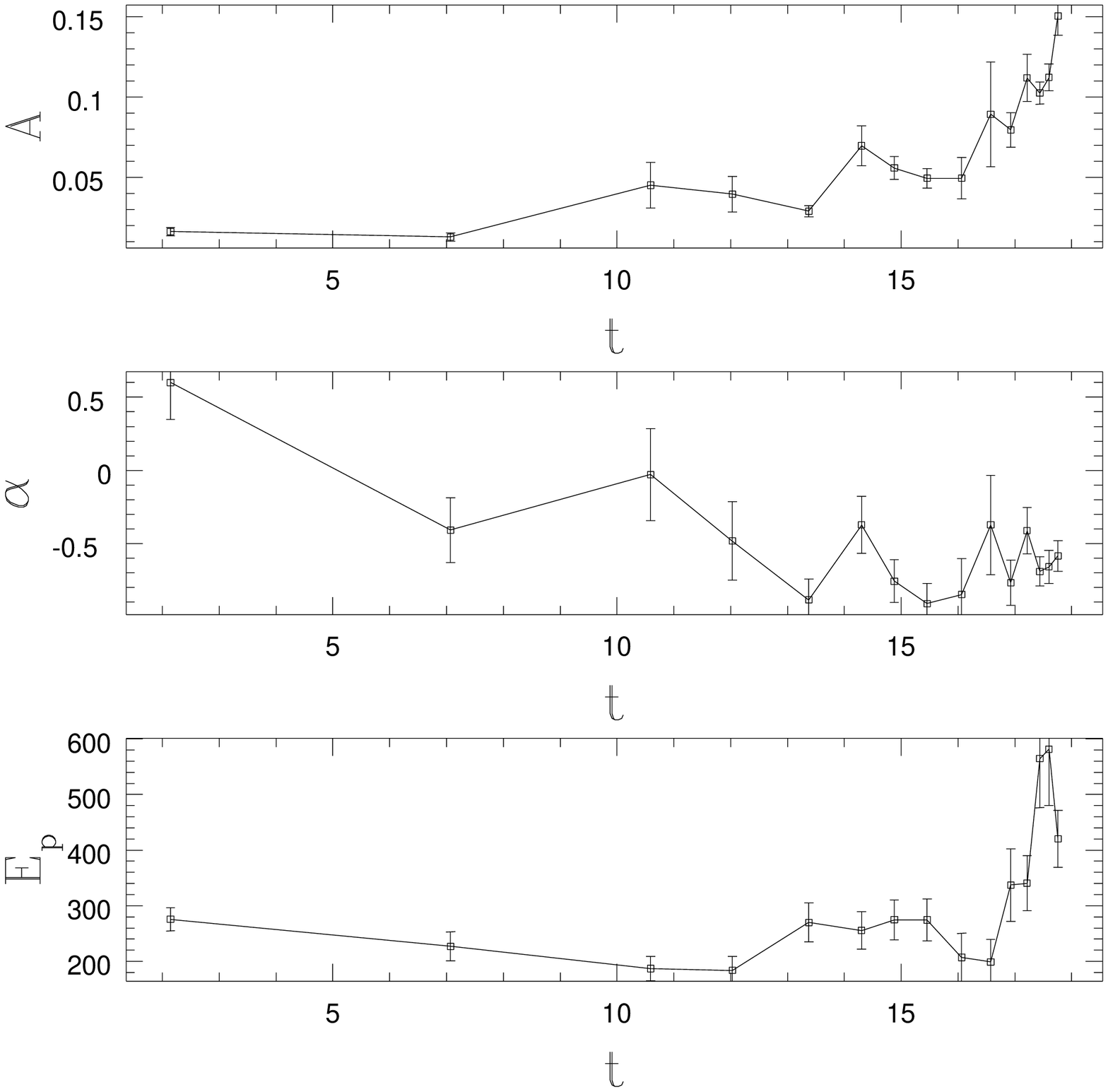,width=0.33\textwidth,height=0.4\textwidth}}
\caption{Three examples of spectral evolution of
$\alpha$, $E_{p}$, and the normalization $A$ 
discussed in the text ((a), left; (b), middle; (c), right)}.  
\end{figure}

\section*{Conclusions}
We conclude that synchrotron emission from internal shocks can reproduce
both the distribution and temporal evolution of GRB spectral parameters.
 Depending on the conditions at the GRB,
synchrotron spectra can have different values for the
low energy asymptote of its spectrum, and the apparent correlation
between $E_{p}$ and $\alpha$ (which results from fitting over
a finite bandpass)
can explain the peak and dispersion of the $\alpha$ distribution.
In addition, we conclude that the electron energy distribution must
flatten or decline at low energies; otherwise, we would see many more
bursts with $\alpha<1.5$.  Finally, allowing for variation of internal
parameters from pulse to pulse as in an internal shock model, synchrotron
emission is consistent with the variety of spectral evolution see in GRBs.

\end{document}